\documentclass[aps,prd,reprint,superscriptaddress,nofootinbib]{revtex4-2}
\usepackage[utf8]{inputenc}
\usepackage{amsmath}
\usepackage{amsfonts}
\usepackage{amssymb}
\usepackage{graphicx}
\usepackage[breaklinks]{hyperref}
\usepackage[caption=false]{subfig}
\usepackage{comment}
\usepackage{bm}
\usepackage{slashed}
\usepackage{mathtools}
\usepackage{siunitx}
\usepackage{booktabs}
\usepackage[para]{threeparttable}
\usepackage{multirow}
\usepackage{xcolor}
\usepackage{soul}
\usepackage[normalem]{ulem}
\usepackage[noabbrev,capitalize]{cleveref}
\usepackage[cal=boondox]{mathalfa}

\newcommand*{\integrand}[1]{\text{d}#1}
\newcommand{\code}[1]{\texttt{#1}}

\newcommand*{\rEP}[2]{\rE^{#1}\,\rP^{#2}}
\newcommand*{\gpccm}{\gram\per\centi\meter\cubed}
\newcommand*{\rsomething}[1]{\mathcal{r}_{#1}}
\newcommand*{\rE}{\rsomething{L}}
\newcommand*{\rP}{\rsomething{H}}
\newcommand*{\rK}{\rsomething{K}}
\newcommand{\dwd}{DWD}
\newcommand{\dwds}{DWDs}
\newcommand{\snr}{\ensuremath{\text{SNR}}}
\newcommand{\mchirp}{\ensuremath{\mathcal{M}}}

\DeclareSIUnit \atomicunit{a.u.}
\DeclareSIUnit \erg{erg}
\DeclareSIUnit \rydberg{Ry}
\DeclareSIUnit \clight{\text{\ensuremath{c}}}
\DeclareSIUnit \yr{yr}
\DeclareSIUnit \parsec{pc}
\DeclareSIUnit \sol{M_{\odot}}
\DeclareSIUnit \evpcs{\electronvolt\per\clight\squared}
\makeatletter
\@ifpackagelater{siunitx}{2021/04/18}
{\sisetup{
	range-exponents=combine ,
	range-phrase=- ,
	print-unity-mantissa=false,
	separate-uncertainty=true,
	range-units=single,
}}
{\sisetup{
	range-phrase=-	,
	separate-uncertainty=true,
	range-units=single,
}
}
\makeatother

\definecolor{GrassGreen}{rgb}{0.105,0.55,0.105}

\def\mnras{Mon.\ Not.\ R.\ Astron.\ Soc.}

\def\aap{Astron.\ Astrophys.}
\def\apj{Astrophys.\ J.}

\def\pr{Phys. Rev.}
\def\prl{Phys.\ Rev.\ Lett.}
\def\prd{Phys.\ Rev.\ D}
\def\prx{Phys.\ Rev.\ X}

\def\physrep{Phys. Rep.}

\def\jcap{{JCAP\ }}

\def\jpcs{J.\ Phys.\ Conf.\ Ser.}
\def\lrr{Living\ Rev.\ Relativ.}
\def\cqg{Classical\ Quant.\ Grav.}
\def\epja{Eur.\ Phys.\ J.\ A.}
\def\jpg{J.\ Phys.\ G\ Nucl.\ Partic.}

\begin{document}

	\title{Exotic Compact Objects: The Dark White Dwarf}
	
	\date{\today}
	
	\author{Michael Ryan} 
	\email{mzr55@psu.edu}
	\affiliation{Institute for Gravitation and the Cosmos, The Pennsylvania State University, University Park, PA 16802, USA}
	\affiliation{Department of Physics, The Pennsylvania State University, University Park, PA, 16802, USA}
	
	\author{David Radice} 
	\email{dur566@psu.edu}
	\affiliation{Institute for Gravitation and the Cosmos, The Pennsylvania State University, University Park, PA 16802, USA}
	\affiliation{Department of Physics, The Pennsylvania State University, University Park, PA, 16802, USA}
	\affiliation{Department of Astronomy \& Astrophysics, The Pennsylvania State University, University Park, PA, 16802, USA}

\begin{abstract}
	Several dark matter models allow for the intriguing possibility of exotic compact object formation. These objects might have unique characteristics that set them apart from their baryonic counterparts. Furthermore, gravitational wave observations of their mergers may provide the only direct window on a potentially entirely hidden sector. Here we discuss dark white dwarfs, starting with an overview of the microphysical model and analytic scaling relations of macroscopic properties derived from the non-relativistic limit. We use the full relativistic formalism to confirm these scaling relations and demonstrate that dark white dwarfs, if they exist, would have masses and tidal deformabilities that are very different from those of baryonic compact objects. Further, and most importantly, we demonstrate that dark white dwarf mergers would be detectable by current or planned gravitational observatories across several orders of magnitude in the particle-mass parameter space. Lastly, we find universal relations analogous to the compactness-Love and binary Love relations in neutron star literature. Using these results, we show that gravitational wave observations would constrain the properties of the dark matter particles constituting these objects.
\end{abstract}
	

	\maketitle
	
\section{Introduction} \label{sec:intro}
	Current dark matter search techniques focus on two primary channels: large-scale structure constraints (e.g. \cite{Planck2020,Markevitch2004}) and direct and indirect detection experiments (e.g. \cite{Clark2020,luxMDM2020,Berlin2019,Undagoitia2015,Mitsou2015}).  While these have constrained several of the bulk properties of dark matter, i.e. that dark matter is cold, particulate, and effectively collisionless on large scales, the current lack of dark matter detection or production has not helped to narrow the space of models. Likewise, the field of astro-particle indirect detection (e.g. \cite{Guepin2021,Genolini2021,Maity2021}), while showing possible signals of interest \cite{Leane2020}, has also not yet produced results.  On the other hand, the advent of gravitational wave observations has opened a new window on the universe that could illuminate the dark sector in a completely novel manner.
	
	Several promising alternative dark matter models have a ``complex" (two or more massive particles) particle zoo, and dissipative interactions create the potential for gravitationally-bound macroscopic structures. Many of these models even form exotic compact objects, with both ``dark" black holes (a normal black hole formed from dark matter instead of baryonic matter) \cite{Buckley2018,DAmico2017,deLavallaz2010,Kouvaris2011,Kouvaris2018,Shandera2018,Singh2020,Choquette2019} and dark (neutron) stars \cite{Giudice2016,Hippert2021,Kouvaris2015,Maselli2017} having been proposed. The merger of these exotic compact objects with each other, or with astrophysical compact objects could be revealed by gravitational wave detectors, such as LIGO. Alternatively, dark matter capture by ordinary compact objects could result in the formation of compact, dark matter cores in their interior\cite{Leung2013,Tolos2015,Bauswein2020,Dengler2021,Gleason2022,Dasgupta2020}, creating hybrid dark/baryonic objects ranging from planetary to stellar masses.
	
	While dark black holes and neutron stars are the obvious counterparts to ordinary black holes and neutron stars, little attention has been paid to the dark equivalent of the third member of the ordinary compact object trio: the (dark) white dwarf (DWD). In the simplest sense, a white dwarf is a compact object predominantly composed of degenerate light electrons and massive nuclei. In  white dwarfs, a balance is struck between gravitational  and (electron-dominated) fermion degeneracy pressure forces. This balance sets many of their macroscopic properties, like the radius and compactness. Above a certain mass (the well-known Chandrasekhar limit), this balance is broken and no static configurations exist. Likewise, above a certain density (the onset of neutronization) these objects again become dynamically unstable to collapse. Here, we consider the dark matter analogs to white dwarfs. These would be compact objects, predominantly composed of two or more fermion species of dark matter, where the pressure support is primarily provided by fermion degeneracy pressure.  Importantly, in hidden, multi-particle dark matter models that lack ``nuclear" reactions, \dwds{} may be the only sub-Chandrasekhar-mass compact objects that can form. Also note that, with this definition, in the limit where the particle species contribute equally to the pressure and mass (the single particle limit), these objects may be macroscopically indistinguishable from dark neutron stars in models with weak or no nuclear interactions (e.g. \cite{Narain2006,Kouvaris2015}).
	
	The literature on objects fitting the definition above is sparse, with few articles discussing objects that fit all three criteria. For example, Narain et al.\cite{Narain2006} describe a general, particle-mass-independent framework for dark, fermionic compact objects, assuming single species composition, even studying the effects of a generic interaction term.  Gross et al.\cite{Gross2021} studied \dwd{}-like objects as a possible final state of collapsed dark quark pockets (assuming multiple species, but identical masses) in an analogue of primordial black holes, computing their mass, radii, and long-term stability with that method. Using a less general approach, we extend these analyses to a two-particle, fermionic gas with potentially varying masses, and include a brief examination of additional binary properties, like the tidal deformability and potential universal relations. Hippert et al.\cite{Hippert2021} mentions the high plausibility of dark white dwarf formation in the Twin Higgs mirror model, but focuses on neutron stars instead. Brandt et al. \cite{Brandt2018} consider the stability of a multi-species model for what they refer to as a pion star, focusing on using lattice QCD methods to obtain a precise equation of state. Meanwhile, the discussion on dark planets and similar low-mass, multi-component objects(\cite{Leung2013,Tolos2015,Dengler2021}) requires the mixing of ordinary matter with dark matter. Consequently, the \dwd{} space has remained unexplored.
	
	We hasten to mention that, like most dark, non-primordial black hole and neutron star models, these objects do not necessarily constitute the entirety of dark matter. While the formation of these objects is outside the scope of this work, we note that regardless of the pathway, the population must obey the constraints imposed by microlensing and similar primordial black hole and massive, compact halo object searches(e.g. \cite{Carr2020,Katz2020}). These searches impose constraints on the fraction of dark matter contained in compact objects versus all dark matter. For the masses considered here, this corresponds to less than $\mathcal{O}(0.01)$ for subsolar mass \dwds{}, decreasing to $\mathcal{O}(10^{-4})$ for objects above \SI{10}{\sol}. Given that ordinary white dwarfs in the Milky Way correspond to a similar $\mathcal{O}(0.01)$ fraction of the ordinary matter \cite{Napiwotzki2009,McMillan2016}, this constraint seems reasonable for models that predict a dark astrophysical formation like \cite{Hippert2021}.
	
	In the following sections we examine the properties of the most basic  \dwd{} model following our definition above: two particle species forming a compact object, with fermion degeneracy pressure providing the dominant support against gravitational collapse. We start with a discussion of the basic properties of \dwd{} that can be inferred analytically in Section \ref{sec:analytics}, including the calculation of an equation of state and scaling relations for the mass, radius, and compactness in the non-relativistic limit. We then discuss the results of the fully relativistic hydrostatic-equilibrium calculations across the particle-mass parameter space, highlighting four example parameter cases in Section \ref{sec:tov} and examining several of the macroscopic attributes, potential universal relations, and implications for \dwd{} merger observations. Importantly, we demonstrate that \dwd{} mergers should be detectable by current and planned gravitational wave observatories across much of the dark parameter space and observations can be used to constrain the dark microphysics.  Lastly, we conclude in Section \ref{sec:conclusion}.
	
\section{Analytic Scaling Relations} \label{sec:analytics}
	
\subsection{Equation of State} \label{sec:eos}
	We consider a simplified, cold compact object comprised of a cloud of degenerate, fundamental fermionic particles, $L$, and $H$. The particles have masses $m_H$ and $m_L$, defined such that $m_H\ge m_L$, analogous to the Standard Model proton and electron. We will use the notation 
	\begin{align}
		\rP = \frac{m_H}{m_p}, \;\;\text{ and }\;\;	\rE =\frac{m_L}{m_e},
	\end{align}
	throughout this and following sections, where $m_p$ and $m_e$ are the proton and electron masses. We will assume approximately neutral ``charge" in bulk, i.e. an equal numbers of $L$ and $H$ particles.
	
	The basic thermodynamic properties of such a cloud are well known (\cite{Chandrasekhar2010,Shapiro1983}), with the number density, pressure, and energy density of a single fermionic particle $f$ given by: 
	\begin{equation}
		n_f = \frac{8 \pi}{3 h^3}p_{\rm fermi}^3 = \frac{x^3}{3 \pi^2 \lambda_f^3}
		\label{eq:degen_density}
	\end{equation}
	\begin{align}
		P_f &= \frac{8\pi m_f^4 c^5}{3h^3}\int_{0}^{x}\frac{y^4}{(1+y^2)^{1/2}}\integrand{y} \nonumber \\
		&= \frac{m_fc^2}{\lambda_f^3}\Bigl[\frac{1}{8\pi^2}\bigl(x(1+x^2)^{1/2}\left( \frac{2}{3}x^2-1\right) \nonumber\\
		&\qquad +\ln\left(x+(1+x^2)^{1/2}\right)\bigr)\Bigr] \nonumber\\
		&= \frac{m_fc^2}{\lambda_f^3} \phi(x)
		\label{eq:degen_pressure}
	\end{align}
	\begin{align}
		\varepsilon_f &= 4 \pi  c^5 m_f^4 \int_{0}^{x} y^2 \sqrt{y^2+1} \integrand{y} \nonumber\\
		&= \frac{m_fc^2}{\lambda_f^3}\Bigl[\frac{1}{8\pi^2}\bigl(x(1+x^2)^{1/2}\left( 1+2x^2\right) \nonumber\\
		&\qquad -\ln\left(x+(1+x^2)^{1/2}\right)\bigr)\Bigl] \nonumber\\
		&= \frac{m_fc^2}{\lambda_f^3} \chi(x).
		\label{eq:degen_energy}
	\end{align}
	Here, $\lambda_f=\hbar/(m_f c)$ is the Compton wavelength, $x$ is the dimensionless Fermi momentum, defined using Equation \eqref{eq:degen_density}, and $y=(p c)/(m_f c^2)$. The total pressure in a two-component gas is then just $P_{\rm tot}=P_H+P_L$, while the total energy density is $\epsilon_{tot} = \epsilon_H + \epsilon_L$. 
	
	We will assume that interparticle interactions contribute at most a small correction to the energy density and will not include their effects here. This follows the standard white dwarf model, where the electrostatic interaction contributes the dominant term, at least at high densities, with a correction to the pressure on the order of the fine structure constant, $(P+P_{\rm correction})/P\approx\SI{0.4}{\percent}$ for a hydrogen white dwarf \cite{Shapiro1983}.  We have chosen this path to preserve the generalizability of these results, since the type of interaction changes between dark matter models. 
	
\subsection{Polytropic approximation} \label{sec:polytrope}
	Since the full EoS is complicated, especially when considered across the $m_L-m_H$ parameter space, it is convenient to expand the EoS as a power series in $x$, keeping only the dominant term. Then the EoS can be approximated as a polytropic function, $P(x)=K x^{\Gamma/3}$, where $K$ and $\Gamma$ are the polytropic constants. Commonly, this is rewritten in terms of the rest mass density ($\rho_0=\sum m_f n_f \approx m_H n_H \approx m_H n_L$) and the polytropic index $n$, as $P(\rho_0)=K \rho_0^{1+1/n}$. 
	
	The polytropic approximation then falls into one of four limiting cases, depending on whether the particles are highly relativistic ($x_f\gg 1$) or non-relativistic ($x_f\ll 1$), and whether the particle masses are substantially different ($m_L\ll m_H$) or similar ($m_L\approx m_H$). Generally, the heavy particles are non-relativistic except in the similar-mass, relativistic electron limit and the similar mass limit can also be thought of as the single particle limit, approximately obtainable with the replacement $L\rightarrow H$. From Equation \eqref{eq:degen_density}, the relativity condition can be written as a condition on $\rho_{0}$, with the non-relativistic (light particle) limit as $\rho_{0}\ll\SI{e6}{\gpccm}\rEP{3}{}$ and the highly relativistic limit $\rho_{0}\gg\SI{e6}{\gpccm}\rEP{3}{}$. 
	
	Using the notation from above and defining $K_{\rm WD}(n)$ as the $n$-dependent polytropic constant for the ordinary white dwarf, the four cases can be written as 
	\begin{equation}
		P_{\rm tot} \approx \begin{cases}
				\rEP{-1}{-5/3} K_{\rm WD}(\frac{3}{2})\,\rho_{0}^{5/3} & (a) \\
				\rP^{-4/3} K_{\rm WD}(3)\,\rho_{0}^{4/3}& (b)  \\
				2\,\rP^{-8/3}K_{\rm WD}(\frac{3}{2})\,\rho_{0}^{5/3} & (c) \\
				2\,\rP^{-4/3}K_{\rm WD}(3)\,\rho_{0}^{4/3} & (d)
		\end{cases},
		\label{eq:poly_scaling}
	\end{equation}
	with $(a)$ being $\rho_{0}\ll\SI{e6}{\gpccm}\rEP{3}{}$, $m_L\ll m_H$, $(b)$ being $\rho_{0}\gg\SI{e6}{\gpccm}\rEP{3}{}$, $m_L\ll m_H$, $(c)$ being $\rho_{0}\ll\SI{e6}{\gpccm}\rP^{4}$, $m_L\approx m_H$, and $(d)$ being $\rho_{0}\gg\SI{e6}{\gpccm}\rP^{4}$, $m_L\approx m_H$. Of note, the $(a)$ and $(b)$ cases correspond to the ordinary white dwarf when $\rP=\rE=1$.

\subsection{Newtonian Hydrostatic Approximation}
\label{sec:newt_approx}
	Next, we examine the parametric dependencies of the \dwd{} mass, radius, and compactness. This can be accomplished by solving the Newtonian hydrostatic equilibrium equations. Defining $m(r)$ as the total mass contained within radius $r$, $p(r)$ as the net outward pressure, and gravitational constant $G$, we have
	\begin{subequations}
		\begin{align}
			\frac{dp}{dr}&=-\frac{G m }{r^2}\rho(r), \label{eq:newton_p}\\
			\frac{dm}{dr}&=4 \pi r^2\rho(r) .
			\label{eq:newton_m}
		\end{align}
	\end{subequations}
	With the inclusion of a polytropic EoS and the definitions, $\rho_c=\rho(r=0)$, $\rho=\rho_c \theta^n$, $r=a\xi$, and $a=[(n+1)K\rho_c^{1/n-1}/(4\pi G)]^{1/2}$,  Equations \eqref{eq:newton_p} and \eqref{eq:newton_m} can be combined into the Lane-Emden equation,
	\begin{equation}
		\frac{1}{\xi^2}\frac{d}{d\xi}\xi^2 \frac{d \theta}{d\xi} = -\theta^n.
		\label{eq:lane_embden}
	\end{equation}
	Note that the only remaining polytropic parameter is the index; this equation is otherwise independent of the EoS and thus mass dependencies. Numeric integration of Equation \eqref{eq:lane_embden} with the boundary conditions $\theta(0)=1$, $\theta'(0)=0$, gives the point $\theta(\xi_1)=0$, which corresponds to the surface of the object. Undoing the previous transformations provides solutions for the final radius ($R$) and mass ($M$) of the \dwd, 
	\begin{align}
		R_{\rm DWD} &= \left(\frac{(n+1)K}{4 \pi G}\right)^{1/2} \rho_c^{(1-n)/2n}\xi_1 \nonumber\\
		M_{\rm DWD} &=4 \pi \left(\frac{(n+1)K}{4 \pi G}\right)^{3/2} \rho_c^{(3-n)/2n}\xi_1^2|\theta'(\xi_1)| \nonumber\\.
	\end{align}
	These equations can be rewritten in terms of the ordinary white dwarf mass and radius using the density scaling $\rho_{\rm DWD} = \rEP{3}{}\rho_{WD}$ and polytropic constant scaling
	\begin{equation}
		\rsomething{K} = \frac{K_{\rm DWD}(n)}{K_{\rm WD}(n)},
	\end{equation}
	 with $K_{\rm DWD}$ from Equation \eqref{eq:poly_scaling}, giving
	\begin{align}
		R_{\rm DWD} 
		&= \left(\rE^3 \rP\right)^{(1-n)/2 n}\rK^{1/2} R_{\rm WD}(\rho_c) \nonumber\\ 
		&= \rEP{-1}{-1} R_{\rm WD}(\rho_c) \label{eq:radius_equation}\\
		M_{\rm DWD}
		&= \left(\rE^3 \rP\right)^{(3-n)/2 n}\rK^{3/2} M_{\rm WD}(\rho_c) \nonumber\\
		&= \rP^{-2} M_{\rm WD}(\rho_c).
		\label{eq:mass_equation}
	\end{align}

	 Unsurprisingly, we recover the classic Chandrasekhar mass limit scaling, Eq.~(\eqref{eq:mass_equation}), commonly seen in the literature on exotic compact objects (e.g. \cite{Giudice2016, Shandera2018, Singh2020}) as either a lower mass bound, when discussing black holes, or an upper mass bound when discussing other types of objects. Lastly, the compactness is given by
	\begin{align}
		C_{\rm DWD}(\rho_c) = \left.\frac{M(\rho_c)}{R(\rho_c)}\right|_{\rm DWD} 
		&= \frac{\rE}{\rP}C_{\rm WD}(\rho_c).
		\label{eq:rescale_compact}
	\end{align}

	As expected, when $m_L \to m_H$ we recover the single-particle limit, 
	\begin{subequations}
		\label{eq:sp_limit}
		\begin{align}
			R_{\rm DWD} &\propto \rP^{-2} R_{\rm SP}(\rho_c) \label{eq:sp_rad}\\
			M_{\rm DWD} &\propto \rP^{-2} M_{\rm SP}(\rho_c) \label{eq:sp_mass}\\
			C_{\rm DWD} &\propto C_{\rm SP}(\rho_c) \label{eq:sp_comp}
		\end{align}
	\end{subequations}
	with the $m_L^{-2}=m_H^{-2}$ scaling seen in the literature \cite{Reisenegger2015}.
	Note that while \crefrange{eq:sp_rad}{eq:sp_comp} display the scaling for dark neutron stars, whose Fermi pressure and energy density are dominated by the neutron terms, they would only be useful for order of magnitude estimation, because the properties of dark neutron stars, like ordinary neutron stars, are heavily influenced by inter-particle interactions that are lacking in this model(see e.g. \cite{Kouvaris2011,Kouvaris2015,Kouvaris2018,Hippert2021}).
	
\section{Numerical Results} \label{sec:tov}
	Now that we have established approximate scaling relations for \dwds{}, we proceed to compute their properties using a fully relativistic treatment. In particular, the Tolman-Oppenheimer-Volkoff (TOV) equations\cite{Shapiro1983,Reisenegger2015},
	\begin{subequations}
		\begin{align}
			\frac{dp}{dr}&=-\frac{G m}{r^2}\left[\frac{\epsilon}{c^2}\right]\left[1+\frac{p}{\epsilon}\right] \times\nonumber\\
			&\qquad\left[1+\frac{4\pi r^3 p}{m c^2}\right]\left[1-\frac{2 Gm}{c^2 r}\right]^{-1}, \label{eq:tov_p}\\
			\frac{dm}{dr}&=4 \pi r^2\left[\frac{\epsilon}{c^{\textbf{\textcolor{blue}{2}}}}\right] ,
			\label{eq:tov_mass}
		\end{align}
	\end{subequations}
	update the Newtonian hydrostatic equilibrium equations (\crefrange{eq:newton_p}{eq:newton_m}) to add corrections due to general relativity (square brackets), necessary to describe the gravitational field of compact objects. We have suppressed the $r$-dependence of $\epsilon$ and $p$ for clarity. Like the Newtonian approximation, the TOV equation can be non-dimensionalized and solved numerically to obtain the approximate scaling relations from \cref{sec:newt_approx}, given the corresponding polytrope\cite{Tooper1965}. As the matter density in an actual \dwd{} could fall anywhere between the non- and highly-relativistic limiting cases, we need to use the full EoS and so a numerical TOV solver. 	To this aim we use a modified version of the \code{TOVL} solver developed by \cite{Damour:2009vw} and \cite{Bernuzzi:2008fu}. 
	
\begin{figure*}[thbp!]
		\centering
		\subfloat[]{%
			\includegraphics[width=\columnwidth]{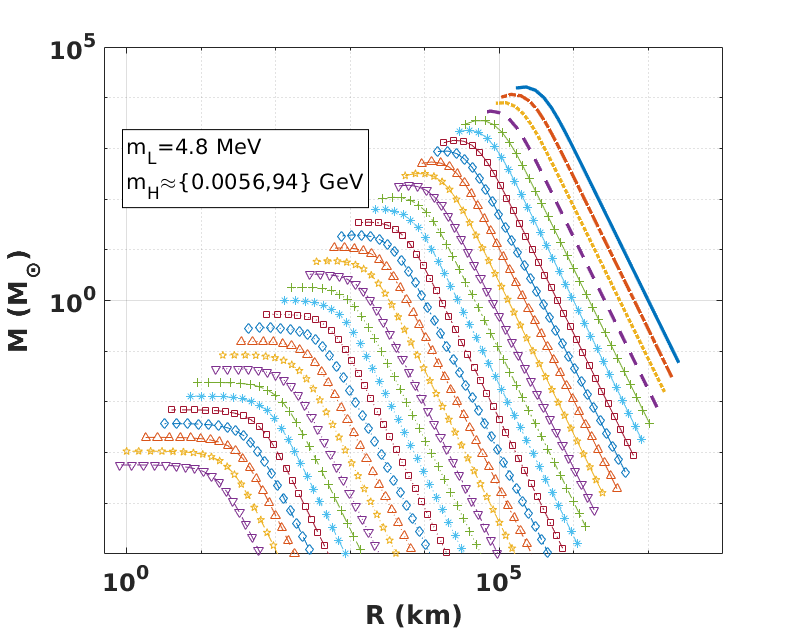}%
			\label{fig:mass_radii_log}
		}
		\subfloat[]{%
			\includegraphics[width=1.05\columnwidth]{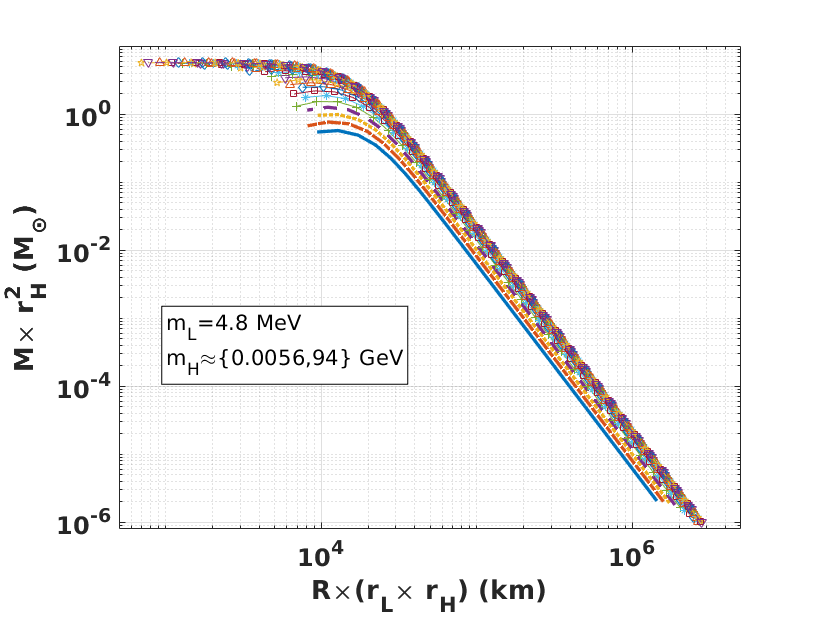}%
			\label{fig:mass_radii_scaled}%
		}
		\caption{
			Example mass-radius relations for varying parameter values. Each line in \cref{fig:mass_radii_log} corresponds to the mass-radius relation for a single value of $m_H$ and the fixed value $m_L=\SI{4.8}{\mega\evpcs}$ resulting from solving the Tolman-Oppenheimer-Volkoff equation over a range of central densities. The blue, solid, markerless line is $m_H=\SI{5.6e-3}{\giga\evpcs}$, with increasing $m_H$ shifting the relation's maximum mass down and to the left.	Note that the maximum mass clearly scales as described in Eq.~(\eqref{eq:mass_equation}), and the $M-R$ curves exhibit the classic white dwarf shape, even when approaching the single-particle limit, $m_H=\SI{5.6e-3}{\giga\evpcs}$ (solid blue, no marker). Above the densities corresponding to the maximum mass (to the left on the plot), the \dwd{} is gravitationally unstable. The cutoffs on the right simply correspond to the minimum density plotted. \Cref{fig:mass_radii_scaled} is identical to \cref{fig:mass_radii_log}, but with radius rescaled as $R\to \rEP{}{}R $ and mass rescaled as $M \to \rP^2  M $.
		}
		\label{fig:mass_radii}
	\end{figure*}

	Solving either the Newtonian approximation or the TOV equation across a range of central densities (i.e. $\rho_c=\rho(r=0)$, as before) for a given $m_L$ and $m_H$ generates a relationship in the $M-R$ space known as a mass-radius relation. In Fig. \ref{fig:mass_radii}, we plot some of these mass-radius relations in the slice of the $m_L-m_H$ parameter space specified by $m_L=\SI{4.1}{\mega\evpcs}$ and $m_H=\SIrange{4.1e-3}{94}{\giga\evpcs}$ on logarithmic $R-M$ axes. Clearly visible are the wide ranges in $M$ and $R$, even over a small range in the $m_H$ parameter space. Conversely, the overall shape of the $M-R$ curve remains similar over that same range. There is a noticeable plateau that appears in the $m_H \gg m_L$ regime and disappears as $m_H \to m_L$. The plateau is due to the fact that the light particle becomes ultrarelativistic in the core of these \dwds{} and the equation of state becomes a polytrope with $\Gamma = 4/3$ (see \cref{sec:polytrope}). As $m_H \to m_L$, the maximum mass is achieved before the particles enter the highly relativistic regime. For example, the maximum mass for $\{m_L,m_H\}=\{4.8,5.6\}\si{\mega\electronvolt}$ (solid blue line) occurs at $x \approx 1.1$, well within the transition regime. The transition to the single particle limit can be seen in the behavior of the radius scaling in \cref{fig:mass_radii_scaled}. As $m_H \to m_L$, the radii begins changing by a factor approaching $2^{2/3}$.  
	
	The dimensionless, gravito-electric quadrupolar tidal deformability ($\Lambda_2$;\cite{Binnington2009}) is also of interest for comparison with gravitational wave observations. The calculation of $\Lambda_{2}$ requires the numeric solution of the reduced, relativistic, quadrupole gravitational potential ($y$) differential equation from \cite{Hinderer2007,Damour:2009vw,Lindblom2014}
	\begin{align}
		\frac{dy}{dr} &= -\frac{y^2}{r} - \frac{r+4\pi r^3(p-\epsilon)}{r(r-2m)}y + \frac{4(m+4\pi r^3 p)^2}{r(r-2m)^2} \nonumber\\
		&\qquad +\frac{6}{r-2m} - \frac{4 \pi r^2}{r-2m}\left[5 \epsilon+9 p+\frac{(\epsilon + p)}{(dp/d\epsilon)^2}\right]
		\label{eq:lindblom_version}
	\end{align}
	This is solved in parallel with the TOV equation to find the value of $y$ at the surface of the object, $Y=y(R)$. 
	
	The quantity $\Lambda_{2}$ is then defined by
	\begin{equation}
		\Lambda_{2}=\frac{2}{3}\frac{k_2}{C^5}
		\label{eq:lambda2_def}
	\end{equation}
	where $k_2$ is the tidal apsidal constant, 
	\begin{widetext}
		\begin{align}
			k_2 &= \frac{8 C^5}{5} \left[(1-2C)^2 (2 C (Y-1)-Y+2)\right] \times \nonumber\\
			&\qquad \Bigl[6C (2 - Y + C(5Y - 8)) + 4C^3[13 - 11Y + C(3Y - 2) + 2C^2 (1 + Y)] \nonumber\\
			&\qquad\qquad  + 3(1 - 2C)^2 [2 - Y + 2C(Y - 1)] \log(1 - 2C)\Bigr]^{-1}.
			\label{eq:k2_full}
		\end{align}
	\end{widetext}
	The $\Lambda_{2}$ parameter specifies how much the object deforms in the tidal field of a companion star and is directly related to the compactness. Black holes, for example, have $\Lambda_{2}=0$, LIGO-detectable neutron stars in binaries are in the range \numrange{1}{e4}, and white dwarfs are \num{>e10}\cite{Binnington2009,Godzieba2020}. The usage of $\Lambda_{2}$ in regards to observations is explained in section \ref{sec:universal}.
	
	Note that we have included two modifications to the \code{TOVL} solver from \cite{Damour:2009vw,Bernuzzi:2008fu}. First, the \code{TOVL} solver, as written, computes the solution to the second-order Regge-Wheeler equation, instead of \cref{eq:lindblom_version}. As the numerical solver had difficulty converging on a solution in parts of the parameter space, and for improved numerical efficiency, we use the equivalent, first-order  \cref{eq:lindblom_version} instead\cite{Hinderer2007,Lindblom2014}. Second, \cref{eq:k2_full} runs into numerical precision difficulties when the compactness is small, with, for instance, terms in the denominator (not) canceling as they should, leading to negative values of $k_2$. We replaced the analytic form of \cref{eq:k2_full} with a series expansion around both $C=0$ and $Y=1$ out to fifth order. This introduces an error of $<1\%$ into the calculations across the entire parameter space defined below.

\subsection{Parameter Space} \label{sec:parameter}
\begin{figure*}[htbp!]
		\centering
		\subfloat[Example Cases]{%
			\includegraphics[width=\columnwidth]{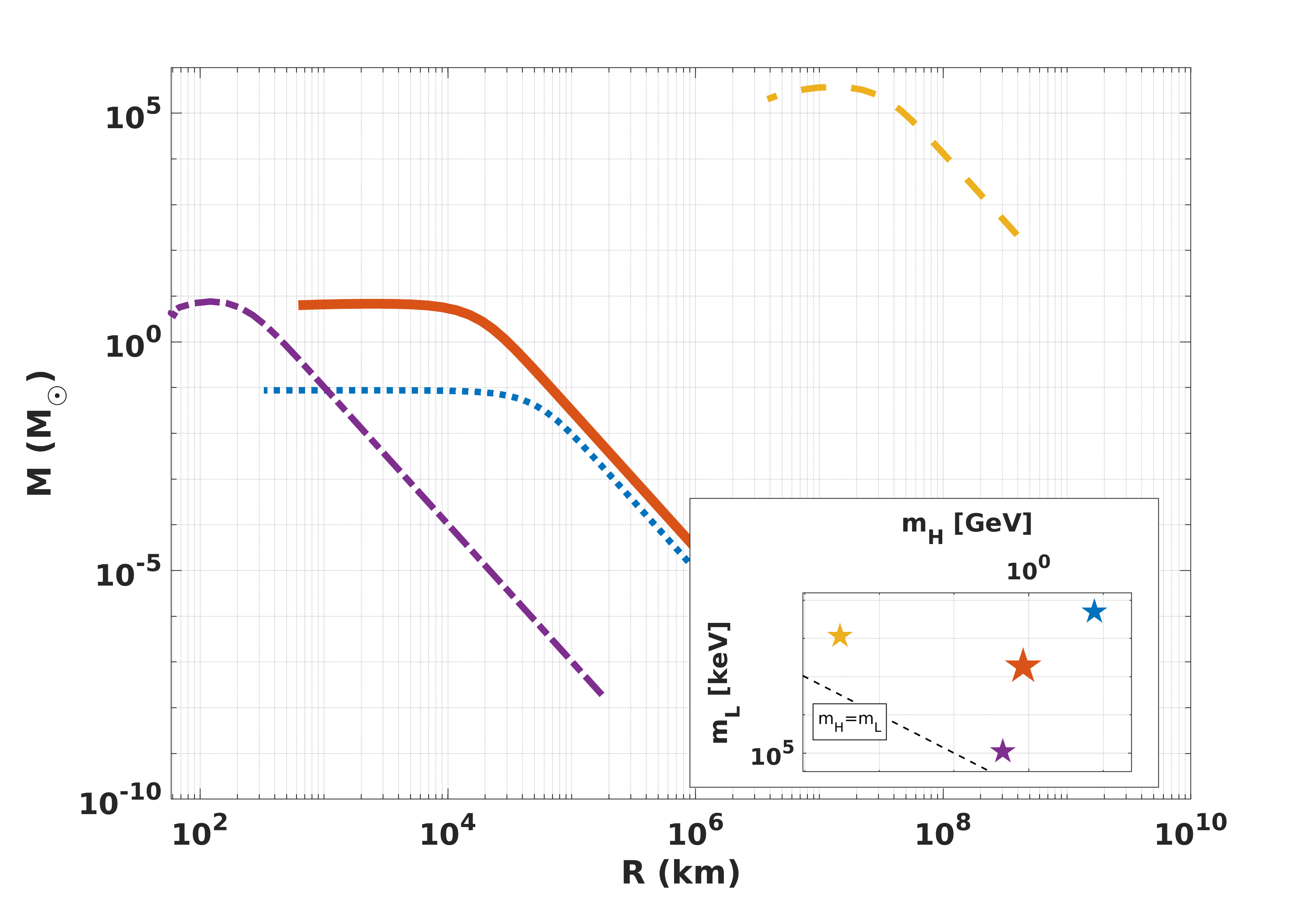}%
			\label{fig:case_study}
		}
		\subfloat[$M_{\rm max}$ in \si{\sol}]{%
			\includegraphics[width=1.05\columnwidth]{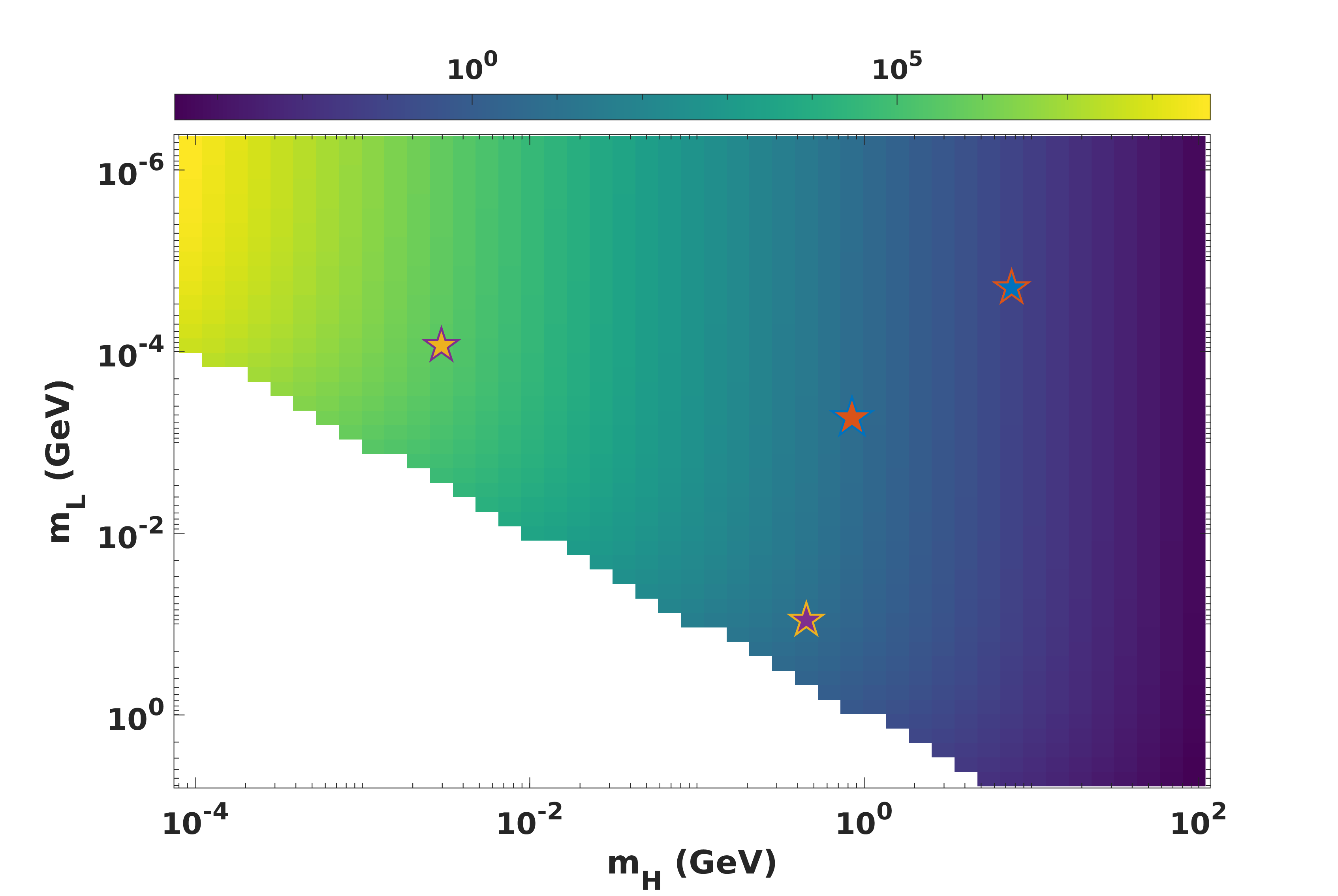}%
			\label{fig:m_max}%
		}
	
		\subfloat[Compactness at $M_{\rm max}$]{%
			\includegraphics[width=1.05\columnwidth]{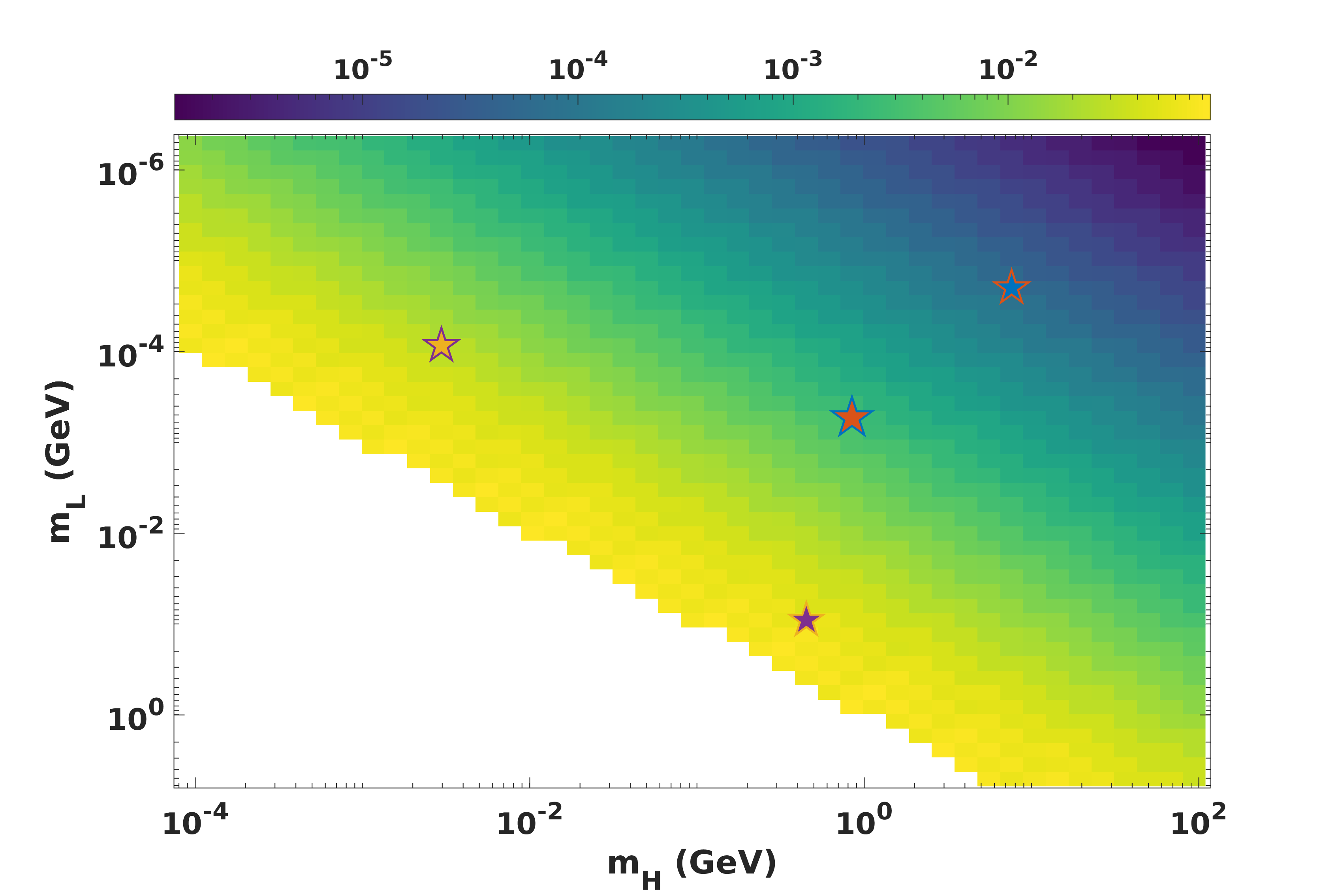}%
			\label{fig:compactness}%
		}
		\subfloat[$\Lambda_2$ at $M_{\rm max}$]{%
			\includegraphics[width=1.05\columnwidth]{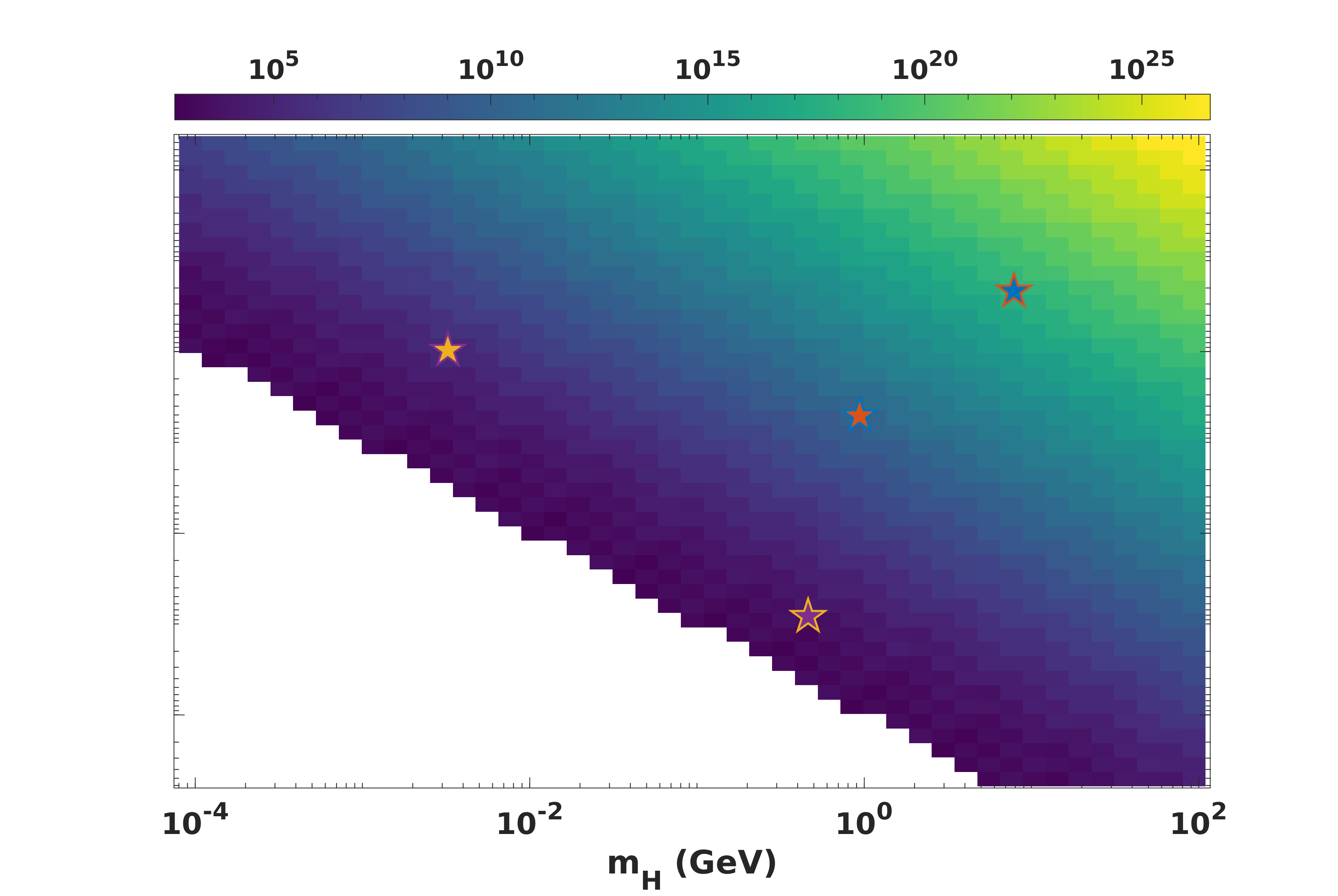}%
			\label{fig:lambda}%
		}		
		\caption{Tollman-Oppenheimer-Volkoff and electric quadrupolar tidal deformability ($\Lambda_2$) solution results for the parameter range $m_L=$ \SI{0.511}{\kilo\evpcs} to \SI{5.11}{\giga\evpcs} and $m_H=$ \SI{93.8}{\kilo\evpcs} to \SI{93.8}{\giga\evpcs}, with values for $m_L>m_H$ ignored. In Fig. \ref{fig:case_study}, we display the mass-radius relations near the three ``corners" of the $m_H-m_L$ parameter space (with corresponding points in the parameter space subfigure) as well as (approximately) the Standard Model white dwarf (thick, solid/largest). Notably, the masses and radii of these objects span many orders of magnitude. Panels (a)-(d) show the maximum mass, compactness and tidal deformability found. The $C$ and $\Lambda_{2}$ values were plotted at the central density corresponding to the maximum mass achieved at that value of $(m_H,m_L)$, and are the maximum (minimum) possible compactness (tidal deformability) for that parameter set.}
		\label{fig:single_dwd}
	\end{figure*}

	In \cref{fig:single_dwd}, we plot the dark white dwarf mass, compactness, and tidal deformability computed by solving the TOV and $\Lambda_2$ equations across the values $m_L=$ \SI{0.511}{\kilo\evpcs} to \SI{5.11}{\giga\evpcs} and $m_H=$ \SI{93.8}{\kilo\evpcs} to \SI{93.8}{\giga\evpcs} (corresponding to $\rE=$ \numrange{e-3}{e4}, $\rP=$ \numrange{e-4}{e2}), with the restriction $m_L\le m_H$ (the behavior is symmetric across the $m_L=m_H$ line). At each sampled $m_H-m_L$ point in the parameter space, the TOV and $\Lambda_2$ equations are solved for $\rho_c$ ranging from \SIrange{e-5}{e25}{\gpccm}. Figure \ref{fig:case_study} shows the mass-radius relations for three parameter cases, somewhat representative of the three parameter space corners: light-light (yellow), heavy-light (blue), and heavy-heavy (purple), and the Standard Model (SM) relation (red), where light (heavy) corresponds to significantly below (above) the Standard Model value.  In Fig. \ref{fig:m_max}, we plot the maximum mass obtained for each mass-radius relation, while in Fig.s \ref{fig:compactness}-\ref{fig:lambda}, we plot $C$ and $\Lambda_2$ at the value of $\rho_c$ corresponding to said maximum mass.
	
	Of note, the maximum mass scales predominantly with $m_H$, as seen by comparing the light-light and SM cases and SM and heavy-heavy cases, and ranges from \SIrange{3e-4}{5e8}{\sol}. The compactness also scales approximately with the ratio $m_L/m_H$, as shown by comparing the light-light and heavy-heavy cases and predicted in the Newtonian approximation, ranging from \numrange{e-6}{0.09}. The scaling with the ratio generally holds for $\Lambda_{2}$ as well, due to the strong dependence on $C$, leading to a minimum of $\sim\num{500}$ near the $m_L\approx m_H$ line and a maximum of \num{4e26} in the upper right corner. The maximum mass configuration is attained at a density that depends on the values of $m_H$ and $m_L$, corresponding to a central Fermi momentum of $x \sim m_H/m_L$, rather than a fixed value. This additional factor corresponds to a change in the central density scaling used in \crefrange{eq:radius_equation}{eq:rescale_compact} from $\rho_c \propto r_L^3 r_H$ to $\rho_c\propto r_L^2 r_H^2$. Substitution  gives $C_{\rm DWD}\propto\left(\rE/\rP\right)^{2/3}$, which is what we see in \cref{fig:compactness} for $m_L\ll m_H$. For $m_L\to m_H$, $C$ approaches the single particle limit, $C=0.114$\cite{Narain2006}.

\subsection{Implications for Gravitational Wave Observations} \label{sec:gwobs}
	In many of the dark matter models the dark sector is mostly or entirely hidden, only observable through gravitational interactions. Thus, \dwd{} observations may be limited to purely gravitational techniques, like, for example, the detection of gravitational waves from the merger of a \dwd{} and some other compact object. As a first step, and since the  \dwds{} span a large range in both mass and compactness, it is worthwhile to determine the detectability across the microphysical parameter space. A gravitational wave observation is detected if the signal-to-noise ratio ($\snr$), defined as\cite{Jaranowski2007}
	\begin{equation}
		\langle\snr^2\rangle =   4 \int_{0}^{\infty}\frac{|\tilde{h}(f)|^2}{S_n(f)}\integrand{f},
		\label{eq:detect_expectation}
	\end{equation}
	for a signal with strain $h(t)$ and Fourier transform $\tilde{h}(f)$ observed by detector with sensitivity curve $S_n(f)$, achieves a specified detection threshold.  As the choice of threshold is somewhat arbitrary, $\snr\ge8$ is used here to match recent LIGO usage\cite{Abbott2020,Abbott2020a}.

\begin{figure}[htbp!]
		\includegraphics[width=\linewidth]{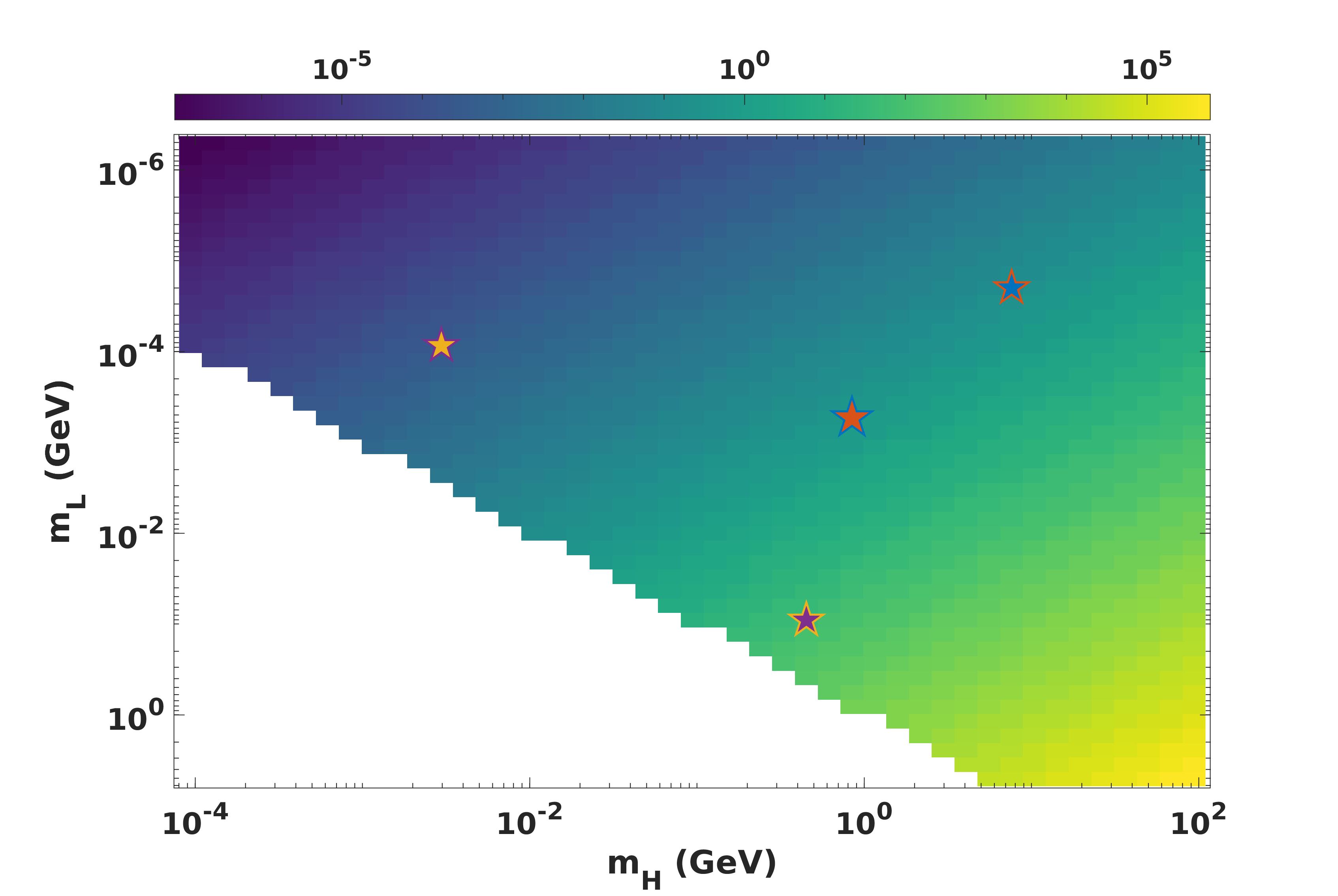}
		\caption{Frequency at merger contact in \si{\hertz} for two identical, maximum-mass dark white dwarfs as a function of $m_H$ and $m_L$.}
		\label{fig:fcontact}
	\end{figure}

	With the assumption that the strain can be approximated as originated from a quadrupole source and truncated to Newtonian order, its Fourier transform can be written as 
	\begin{equation}
		\tilde{h}(f) \approx \frac{\sqrt{5/24}}{\pi^{2/3}D_L} M_C^{5/6} f^{-7/6},
		\label{eq:strain_fourier}
	\end{equation}
	where $M_C= (M_1 M_2)^{3/5}/(M_1+M_2)^{1/5}$ is the chirp mass and $D_L$ is the luminosity distance of the merger. 
	We will restrict the analysis to the in-spiral phase of the merger; postmerger components, especially for high mass objects, will require numerical relativity simulations. This reduces the integral bounds to $0<f<f_{\rm contact}$, where $f_{\rm contact}$ is the contact frequency, i.e. the binary orbital frequency at the termination of the in-spiral period.  Using $f_{\rm contact}$ will possibly overestimate the final $\snr$, since it does not, for example, account for tidal effects (see, e.g., \cite{DeLuca2021} for other choices), but it does provide a reasonable, simple estimate\cite{Bernuzzi2012}. In Fig. \ref{fig:fcontact}, we consider two identical, maximum-mass \dwds{}, and plot the contact frequency given by\cite{Damour:2009vw,Bernuzzi:2008fu} 
	\begin{align}
		f_{\rm contact} 
		&= \sqrt{\frac{G}{4 \pi ^2}\frac{2 M_{\rm max}}{(2 R_{M_{\rm max}})^3}}.
		\label{eq:f_contact}
	\end{align}
	Using identical, maximum-mass \dwds{} to calculate $f_{\rm contact}$ provides an optimistic estimate of the maximum frequency emitted during the merger. As real \dwds{} are not likely to be at the maximum mass, actual contact frequencies will be lower. From \cref{fig:fcontact}, we can see that the contact frequency in the optimistic case ranges from \SI{8e-8}{\hertz}, for the most massive \dwds{} in the top right corner, to \SI{600}{\kilo\hertz}, for the least massive \dwds{} in the bottom left. 
	
\begin{figure}[htbp!]
		\includegraphics[width=\columnwidth]{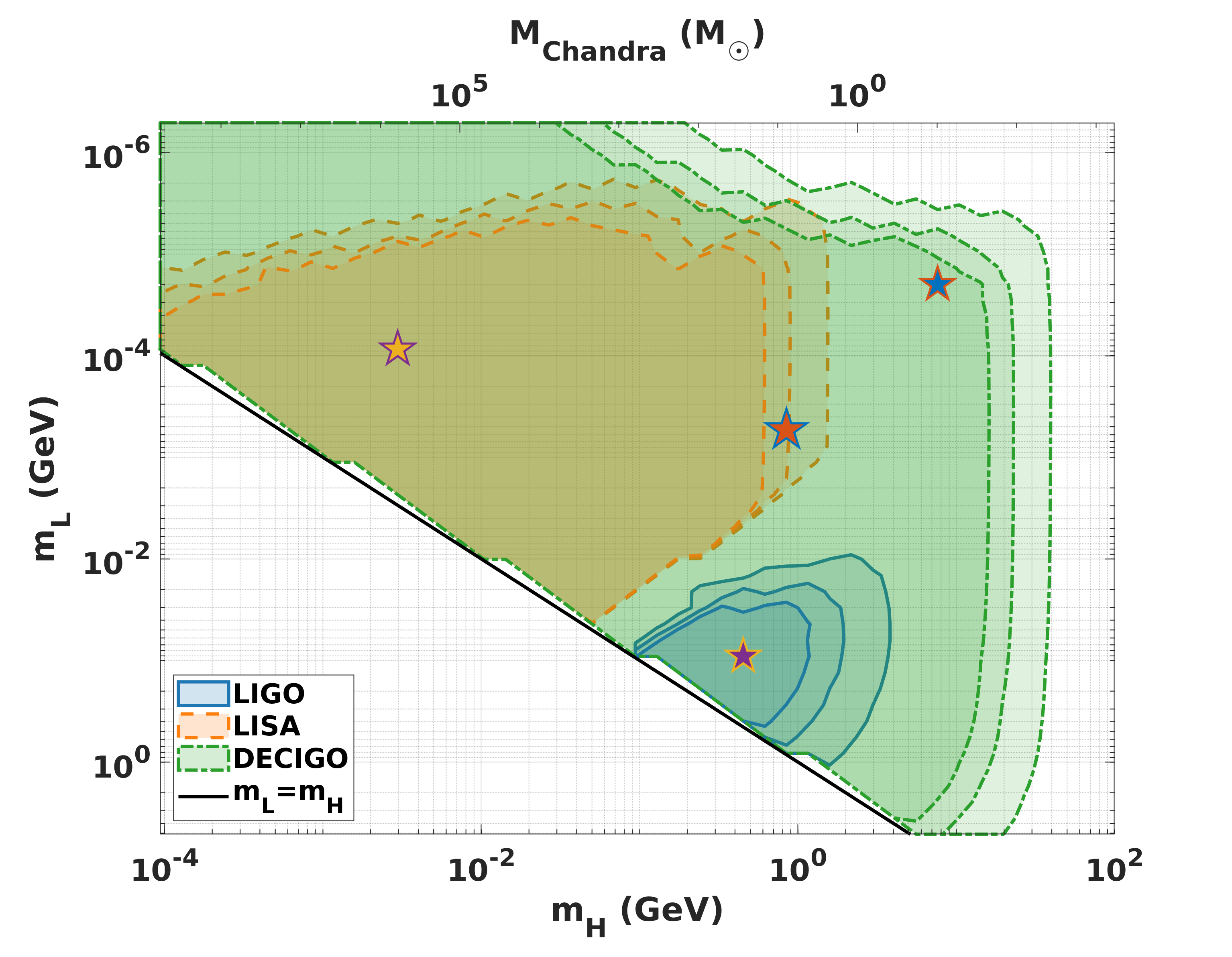} 
		\caption{Gravitational wave detectability ($\snr$) of identical, maximum-mass  \dwd{} mergers. Shaded regions correspond to $\snr\ge8$.  $\snr$ contours are derived from Eqs.~(\eqref{eq:detect_expectation}) and (\eqref{eq:strain_fourier}) with a frequency cutoff of $f_{\rm contact}$ (Eq. \eqref{eq:f_contact}) and plotted at three luminosity distances, $D_L=\{100,250,450\}\si{\mega\parsec}$ (lighter to darker) using design sensitivity curves for LIGO, LISA and DECIGO \cite{LIGO2018,Robson2018,Kawamura2008}. Clearly, the different detectors will probe different regions of the parameter space in a complementary fashion and  combined, LIGO and LISA will probe much of the $m_L\sim\SIrange{e-3}{1}{\giga\electronvolt}$, $m_H\sim\SIrange{e-4}{3}{\giga\electronvolt}$ parameter space. 
		}
		\label{fig:gwo_detect}
	\end{figure}

	Figure \ref{fig:gwo_detect} demonstrates the application of \cref{eq:detect_expectation,eq:strain_fourier,eq:f_contact} using the current sensitivity curves for Advanced LIGO, and the design sensitivity curves for LISA and DECIGO \cite{LIGO2018,Robson2018,Kawamura2008}. We compute the $\snr$ at a luminosity distance of $\{100, 250, 450\}\si{\mega\parsec}$, multiplied by a factor $4/25$ to include an averaging over sky position, inclination, and polarization and assuming a source dominated by quadrupolar radiation \cite{Dominik2014}, and shade the region satisfying the condition $\snr>8$. The contact frequency was computed in the same manner as in Fig. \ref{fig:fcontact}, i.e. assuming two, identical, maximum-mass \dwds{}. As hinted at in Fig. \ref{fig:fcontact}, since the different gravitational wave observatories are sensitive over different frequency ranges, the different parameter cases will be visible by different observatories. While LIGO can observe only the $m_L\in\SIrange{0.01}{1}{\giga\electronvolt}$, $m_H\in\SIrange{.1}{3}{\giga\electronvolt}$ region, corresponding to ordinary-neutron-star-like \dwds{} and the heavy-heavy case, LISA and especially DECIGO should be able to explore a much wider range of parameter space. LISA can probe the $m_L\sim\SIrange{e-6}{0.01}{\giga\electronvolt}$, $m_H\sim\SIrange{e-4}{1}{\giga\electronvolt}$ regime, encompassing the light-light and SM cases, and nicely complimenting the LIGO region. Finally DECIGO would be able to explore below \SI{e-6}{\giga\electronvolt}(\SI{e-4}{\giga\electronvolt}) and up to \SI{1}{\giga\electronvolt}(\SI{30}{\giga\electronvolt}) in $m_L$($m_H$) space, verifying LIGO/LISA results and even including the light-heavy cases.

\subsection{Universal Relations} \label{sec:universal}
	Potential universal relations, especially those of the electric quadrupolar tidal deformability, are of further interest for potential gravitational wave observations. A universal relation is a relation between two or more macrophysical properties that is generally independent of the equation of state, and, more importantly in our case, allows the breaking of observational degeneracies. Consider an example \dwd{} merger, with \dwd{} 1 having macroscopic parameters $\{M_1,C_1,\Lambda_{2,1}\}$ and \dwd{} 2 having $\{M_2,C_2,\Lambda_{2,2}\}$ (assume $M_2<M_1$). The corresponding gravitational wave detection would observe the chirp mass, $\mchirp=(M_1 M_2)^{3/5}/(M_1+M_2)^{1/5}$, mass ratio, $q=M_2/M_1$, and reduced tidal deformability  \cite{Favata2013},
	\begin{align}
		\tilde{\Lambda} &= \frac{16}{13}\frac{1}{(1+q)^5} \Bigl[\bigl(q^4(q+12) - (1+12q)\bigr)\Lambda_A \nonumber\\
		&\qquad + \bigl(q^4(12+q)+(1+12q)\bigr)\Lambda_S\Bigr],
		\label{eq:reduced_tidal}
	\end{align}
	where $\Lambda_S$ and $\Lambda_A$ are the symmetric and anti-symmetric tidal deformabilities,
	\begin{equation}
		\Lambda_{S,A} = \frac{1}{2}(\Lambda_{2,2}\pm\Lambda_{2,1}).
	\end{equation} 

	While the mass ratio can be measured directly from the gravitational wave signal, the tidal deformability enters at leading order in the phasing only through $\tilde\Lambda$. Further, the radii and compactness of the two objects do not directly enter into the phasing or magnitude of the signal. This is where universal relations are useful: breaking the $\Lambda_A,\Lambda_S$ degeneracy in \cref{eq:reduced_tidal} and calculating $C$ and $R$ from the $\Lambda_{2}$ of the individual \dwds{}. First, the binary love relation, discovered by Yagi and Yunes in 2015 while examining binary neutron star properties \cite{Yagi2015}, with the form
	\begin{align}
		\Lambda_{A}(q,\Lambda_S) &= F_n(q) \frac{1 + \sum_{i=1}^{3} \sum_{j=1}^{2} b_{ij} q^j \Lambda_S^{i/5}}{1 + \sum_{i=1}^{3} \sum_{j=1}^{2} c_{ij} q^j \Lambda_S^{i/5}}\Lambda_S
		\label{eq:lambdaa_fit}
	\end{align}
	where $b_{ij}$ and $c_{ij}$ are numerical fitting coefficients and $F_n(q)=(1-q^{10/(3-n)})/(1+q^{10/(3-n)})$ is a polytropic-index-dependent controlling factor, lets Equation \eqref{eq:reduced_tidal} be rewritten as a function of $q$ and $\Lambda_S$ only. From this, one can solve for $\Lambda_S$ and then $\Lambda_A$.	Then the individual $\Lambda_{2,1},\Lambda_{2,2}$ can be computed using the definitions of $\Lambda_S,\Lambda_A$. 
	
	Second, in 2013, Yagi and Yunes \cite{Yagi2013} and Maselli et al.\cite{Maselli2013} demonstrated that the relation between $C$ and $\Lambda_{2}$ was also universal. This relation, which follows mostly from Equation \eqref{eq:lambda2_def} provides an estimate of $C_1$ and $C_2$ from $\Lambda_{2,1}$ and $\Lambda_{2,2}$. The radii then follow directly from the definition of $C$.
	
\begin{figure*}[htbp!]
	\subfloat[$\Lambda_A$ vs $\Lambda_{A,{\rm fit}}$]{
		\includegraphics[width=0.49\linewidth]{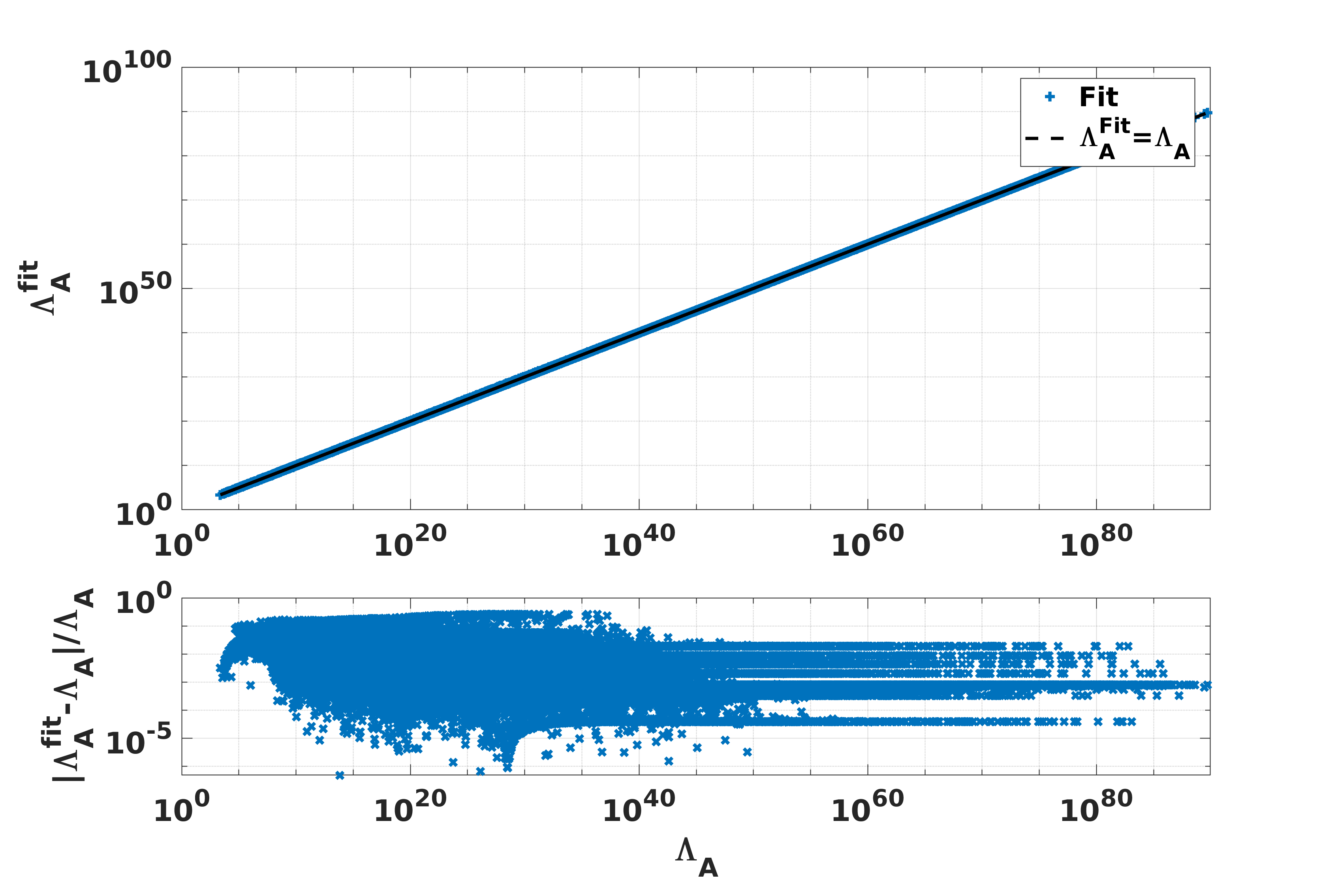}
		\label{fig:lambdaA_results}
	}
	\subfloat[$\Lambda$ vs C]{
		\includegraphics[width=0.49\linewidth]{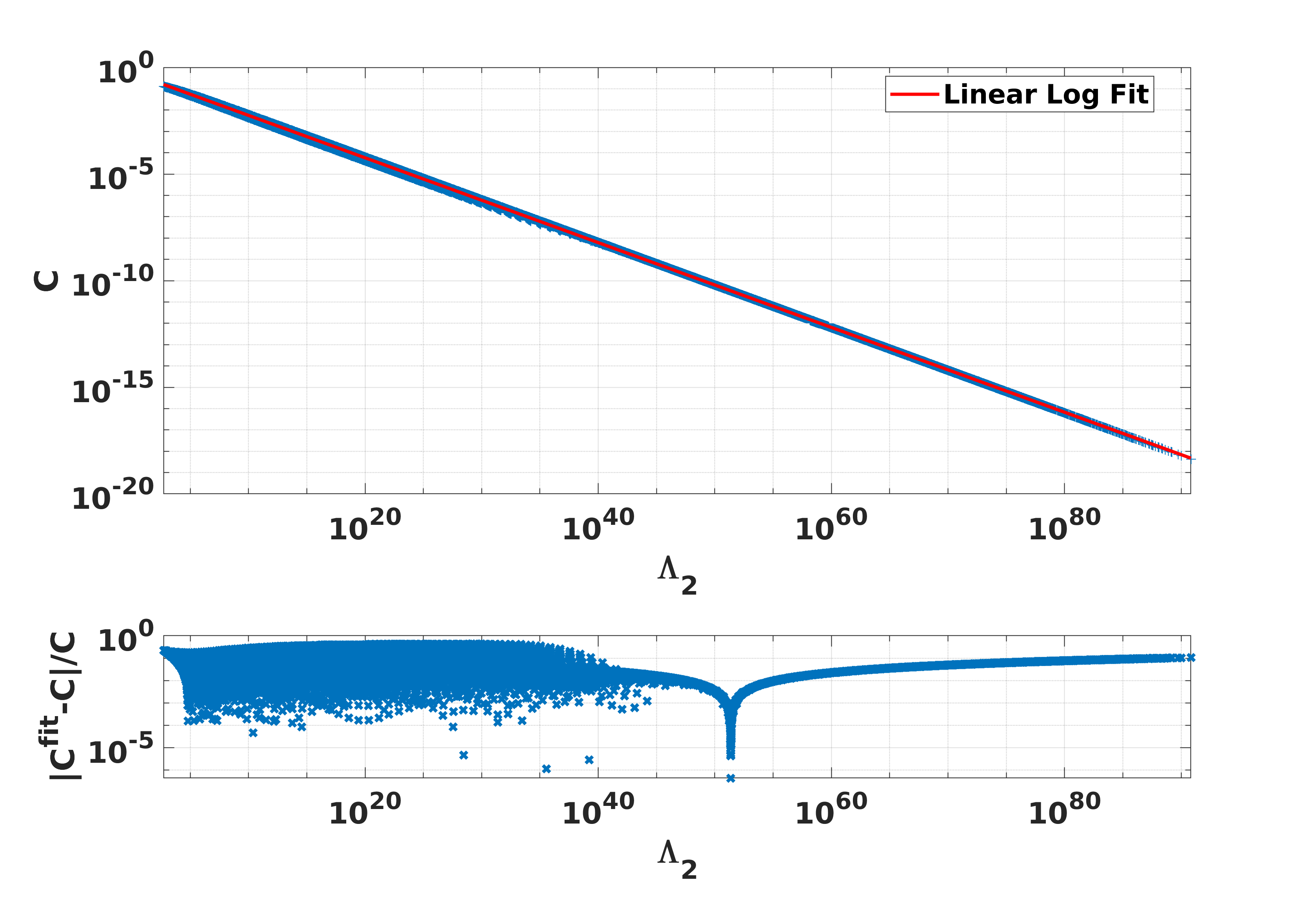}
		\label{fig:lambdaC_results}
	}
	\caption{\dwd{} universal relations. The first panel shows that the binary Love relation, $\Lambda_A=\Lambda_A(q,\Lambda_S)$, is reasonably approximated by a functional form from the neutron star literature \cite{Yagi2015,Yagi2016,Godzieba2020}, using the new coefficients from Table \ref{tab:lambdaa_fit}, whereas the second panel demonstrates that for \dwds{} $\Lambda_2$ can be well approximated by the fit given in Equation \eqref{eq:lambdaC_fit}, effectively $\Lambda_2 \propto C^{5.1}$.}
	\label{fig:universal}
\end{figure*}

	In Fig. \ref{fig:lambdaA_results}, we demonstrate that the $\Lambda_A(q,\Lambda_S)$ function from \cref{eq:lambdaa_fit} also applies to \dwds{}. Here, we considered 2025 $m_L{-}m_H$ pairs in the ranges $m_L\in\{\num{5.11e-7},\num{5.11}\}\si{\giga\electronvolt}$ and $m_H\in\{\num{9.38e-3},93.8\}\si{\giga\electronvolt}$. For each $m_L{-}m_H$ pair, we picked 20 random pairs of central densities and computed $\{M_1,M_2,\Lambda_{2,1},\Lambda_{2,2}, q, \Lambda_S, \Lambda_A\}$, as well as $\Lambda_{A,{\rm fit}}$. The fit is computed using the functional form from \cref{eq:lambdaa_fit} with new $b_{ij}$ and $c_{ij}$ coefficients given in Table \ref{tab:lambdaa_fit}. Note that for our basic model the $F_n(q)$ controlling factor is $\approx 1$ and has been dropped.
	\begin{table}
		\scriptsize
		\centering
		\begin{tabular}{c c c c c c c}
			\toprule
			& $\square_{1,1}$ & $\square_{1,2}$ & $\square_{2,1}$ &$\square_{2,2}$ & $\square_{3,1}$ & $\square_{3,2}$   \\
			\midrule
			$b_{ij}$ & \num{1.73e+00} &\num{-1.57e+00} &\num{5.48e-02} &\num{-5.10e-02} &\num{1.27e-06} &\num{-7.15e-07} \\
			$c_{ij}$ & \num{1.68e+00} &\num{-1.42e+00} &\num{5.47e-02} &\num{-5.01e-02} &\num{1.27e-06} &\num{-6.80e-07} \\
			\bottomrule
		\end{tabular}
		\caption{Fitting coefficients for the binary love relation given in Equation \eqref{eq:lambdaa_fit}.\label{tab:lambdaa_fit}}
		
	\end{table}

	In Fig. \ref{fig:lambdaC_results} we plot $\Lambda_2$ versus $C$ for the 2025 $m_L{-}m_H$ pairs mentioned previously, at logrithmically spaced central densities from \SI{e-5}{\gpccm} to the density corresponding to the maximum mass (approximately \num{56000} \dwds{}). Comparing with the simple linear log fit, 
	\begin{equation}
		\log_{10} C = -0.1958 \log_{10}(\Lambda_2) - 0.3931,
		\label{eq:lambdaC_fit}
	\end{equation}
	which provides an excellent fit over the entire range, we see that $\Lambda_2 \propto C^{-5.1}$ to good approximation. This should not be surprising; even though Equation \eqref{eq:k2_full} appears to show $k_2$ has a strong ($C^5$) dependence on $C$, in reality, the dependence is actually relatively weak,\footnote{The lower order terms in the denominator cancel, leaving $C^5$ as the lowest surviving term. This then cancels with the $C^5$ in the numerator and $k_2$ approaches a nonzero constant as $C\rightarrow0$.} functionally leaving $\Lambda\sim C^{-5}$. It is important to point out this relationship is effectively independent of the dark parameters for this simple model and thus also ``universal".
	
	With these relations, we can substitute some numbers and see what sort of constraints we might be able to place on the dark parameters from an observation. For example, let us consider an hypothetical scenario in which a binary \dwd{} is detected with $q=\num{0.705\pm0.04}$, $\mchirp= \SI{19\pm4}{\sol}$, and $\tilde{\Lambda}=\num{9.09\pm0.9e4}$. Using Equations \eqref{eq:reduced_tidal} and \eqref{eq:lambdaa_fit}, and simply propagating the bounds, we would obtain $\Lambda_S=\num{1.6\pm0.3e5}$ and $\Lambda_A=1.5_{-0.3}^{+0.4}\times10^{5}$. By definition, we would then have $\Lambda_{2,1}=9.52_{-10}^{+6}\times10^{3}$ and $\Lambda_{2,2}=3.1_{-0.6}^{+0.8}\times10^{5}$, which, using our $C-\Lambda_{2}$ relation would give $C_1=\num{0.0887\pm0.02}$ and $C_2=0.0443_{-0.009}^{+0.01}$. Using $m_1 =q^{-3/5} (1 + q)^{1/5} \mchirp $ and $m_2 = q^{2/5} (1 + q)^{1/5}\mchirp$, the chirp mass and mass ratio would give $m_1=\SI{26\pm6}{\sol}$ and $m_2=\SI{18.3\pm4}{\sol}$. From this, the maximum mass would be at least \SI{20}{\sol}, and we could constrain the heavy particle mass, $m_H<\SI{0.45}{\giga\evpcs}$. Further, while the conservative compactness of $C_{max}<0.1087$ would provide a minimal constraint on the heavy to light particle mass ratio ($m_H/m_L>2$), the other end would suggest a not dissimilar constraint of $m_H/m_L>5$, so $m_L$ would likely be \SIrange{90}{200}{\mega\evpcs} (assuming $m_H=\SI{0.45}{\giga\evpcs}$).

\section{Conclusion} \label{sec:conclusion}
We present a first look at the \dwd{}, using a basic model of two, different-mass fundamental fermions to explore some of the possibilities of this exotic compact object. We determine analytic scaling relations for the mass, radius, and compactness of the \dwd{} as function of the Standard Model white dwarf and the fermion masses (Equations \eqref{eq:degen_density}-\eqref{eq:degen_energy} and \eqref{eq:mass_equation}-\eqref{eq:rescale_compact}) in the non-relativistic limit. We accomplish this by solving the Newtonian hydrostatic-equilibrium approximation using the well-known equation of state of fermionic ideal gasses. As expected, we recover the scaling relations found in the literature upon approaching the single-particle limit.

We then solve the Tollman-Oppenheimer-Volkoff and tidal-deformability differential equations numerically to obtain fully relativistic versions of the Newtonian approximations. Using the relativistic formalism confirms the approximate Newtonian scaling as well as highlights the large span of the macrophysical and even binary attributes (Figures \ref{fig:mass_radii}-\ref{fig:fcontact}). 

We further find universal relations between macroscopic properties of \dwds{}  that are analogous to those found for  neutron stars. In particular, we investigate the $C$ vs $\Lambda_2$ and binary love universal relations, with the net result that the $\Lambda_2$-$C$ relationship can be well approximated by a simple power law and that binary love relation can be well approximated by fits from the neutron star literature (Figure \ref{fig:universal}). These relations could be used to determine the radii of \dwd{} from gravitational wave observations of their mergers, thus directly constraining the masses of the dark particles.

Lastly, we discuss the detectability of \dwd{} binary mergers across the fermion mass parameter space. We show that not only are \dwd{} mergers detectable but that, assuming design sensitivity, different gravitational wave observatories would probe different regions of the space (Figure \ref{fig:gwo_detect}). For example, LIGO should be able to detect mergers of high-compactness, lower-mass \dwds{} corresponding to a dark light particle in the mass range \SIrange{0.01}{1}{\giga\electronvolt} and dark heavy particle in the range \SIrange{0.1}{3}{\giga\electronvolt}, while LISA could detect both higher-mass, high-compactness and lower-mass, lower-compactness \dwds{}, corresponding to light particles that are \SIrange{e-6}{0.01}{\giga\electronvolt} and heavy particles that are \SIrange{e-4}{1}{\giga\electronvolt}. Later-generation space-based detectors like DECIGO may be able to detect mergers across an even larger part of the parameter space.

We have left four significant topics to future work, though two of those are interrelated. First, is the effect of inter-particle interactions, both those that do and do not change particle type, on the equation of state. While it is reasonable to ignore particle-conversion interactions, as many dark matter models do not contain them, interactions such as the dark electromagnetism of atomic dark matter\cite{Kaplan2010} or the Yukawa interactions of the model in \cite{Kaplan2009} should also be studied. The lack of such an interaction term is not fatal; after all, the model presented here works quite well for estimating ordinary white dwarf properties and should also work well in cases with weak dark interactions. Conversely, dissipative interactions are the entire point of dissipative dark matter models, necessitating their inclusion in follow-up work. Doing so in a general fashion is non-trivial, but including additional polynomial terms into the total pressure and energy equations as in \cite{Narain2006}, similar to the electrostatic correction in ordinary white dwarfs\cite{Shapiro1983} or the Yukawa term in Kouvaris and Nielsen\cite{Kouvaris2015} would likely be a good first step.

Second, we have restricted our gravitational wave signal analysis to the detectability of the in-spiral portion only. As demonstrated in \cref{fig:lambda}, the tidal deformability of these \dwds{} can be significantly larger than that of ordinary neutron stars and black holes. As such, general template bank searches based on ordinary binary black hole or binary neutron star mergers may not find these objects. Additionally, the post-merger portion of the signal may contain various features that could be strongly dependent on the equation of state and the dark microphysics. Computing the full merger and post-merger signal using numerical relativity simulations at several points in the parameter space would help resolve both of these issues, providing data for both a more targeted search and demonstrating any potential microphysical dependence.

The remaining two issues concern the formation and populations of these objects and using gravitational wave observations to constrain their properties. Just as there are a number of dark matter models that have the particle types to create \dwds{}, so are there a number of possible formation mechanisms, ranging from primordial direct collapse, as in \cite{Gross2021}, to astrophysical direct collapse, analogous to the dark black hole formation or asymmetric stars in e.g. \cite{Buckley2018,Shandera2018,Chang2019}, to the astrophysical remnant of a dark star, suggested in \cite{Hippert2021}. This makes estimating the \dwd{} population highly model-dependent. The possibility that such binaries might not be able to form naturally at all also cannot be excluded. On the other hand, determining the current constraints on the merger rates from LIGO observations should be tractable, assuming the current state of nondetection holds. Likewise, identifying a particular merger as a possible \dwd{} merger (as opposed to an ordinary object merger) should not be difficult, given the significant discrepancies between \dwd{} and ordinary compact object characteristics across the majority of the parameter space. Distinguishing between a \dwd{} and a dark neutron star, or determining the dark composition pose a much higher level of difficulty, however, given the potential overlap in macroscopic traits, and may require population analysis, circling back to the formation problem. 

While there are several major questions left to be resolved, the potential for \dwds{} and their mergers to shine a light on the dark sector strongly motivates the development of targeted search strategies in gravitational wave detectors data.

\acknowledgments
Funding for this work was provided by the Charles E. Kaufman Foundation of the Pittsburgh Foundation. The authors also thank Rahul Kashyap and Daniel Godzieba for their input on the \code{TOVL} and $\Lambda_2$ calculations. 
D.R. acknowledges funding from the U.S. Department of Energy, Office of Science, Division of Nuclear Physics under Award Number(s) DE-SC0021177 and from the National Science Foundation under Grants No. PHY-2011725, PHY-2020275, PHY-2116686, and AST-2108467.
\bibliography{dwd}
	
\appendix

\end{document}